\documentclass[twocolumn]{jetpl}
\usepackage[T1,T2A]{fontenc}
\usepackage{cite}
\usepackage{physics}
\usepackage{mathtools}
\usepackage{xcolor}
\usepackage[russian]{babel}

\rus

\title{Multiqubit GHZ state preparation with $^{171}\text{Yb}^+$ ions for frequency standards}

\rtitle{Подготовка многочастичных GHZ состояний на ионах $^{171}\text{Yb}^+$ для применения в стандартах частоты}

\sodtitle{Подготовка многочастичных GHZ состояний на ионах $^{171}\text{Yb}^+$ для применения в стандартах частоты}

\author{
A.\,E.\,Korolkov$^{1,2}$, I.\,V.\,Zalivako$^{1,2}$, A.\,S.\,Borisenko$^{1,2}$, V.\,N.\,Smirnov$^{1,2}$, P.\,A.\,Kamenskikh$^{1,2}$, I.\,A.\,Semerikov$^{1,2}$, K.\,Yu.\,Khabarova$^{1,2}$, N.\,N.\,Kolachevsky$^{1,2}$
}

\rauthor{Корольков, Заливако, Борисенко, Смирнов, Каменских, Семериков, Хабарова, Колачевский}

\sodauthor{А.\,Е.\,Корольков$^{+*}$\thanks{e-mail: korolkovae@lebedev.ru}, И.\,В.\,Заливако$^{+*}$, А.\,С.\,Борисенко$^{+*}$, В.\,Н.\,Смирнов$^{+*}$, П.\,А.\,Каменских$^{+*}$, И.\,А.\,Семериков$^{+*}$, К.\,Ю.\,Хабарова$^{+*}$, Н.\,Н.\,Колачевский$^{+*}$}

\address{~\\$^1$P.N. Lebedev Physical Institute of the Russian Academy of Sciences, Moscow 119991, Russia\\~\\
$^2$Russian Quantum Center, Skolkovo, Moscow 121205, Russia}

\dates{}{\today}

\abstract {Multiqubit Greenberger-Horne-Zeilinger (GHZ) state preparation is demonstrated on two to eight qubits in the chain of ten $^{171}Yb^+$ ions in the linear Paul trap with a sequence of single-qubit and two-qubit operations and pulsed dynamical decupling. Fidelity of the 8-qubit GHZ state is estimated to be $58.9 \pm 0.6\%$. The expected enhanced sensitivity of the parity oscillations phase to the analyzing pulse phase is observed. This result is a step towards a multi-ion optical clock with ytterbium ions with frequency averaging following $1/N$ scaling law in contrast to the slower $1/\sqrt{N}$ scaling for an ensemble of independent particles in case when the decoherence is dominated by the spontaneous decay.
} 


\begin{document}

\maketitle

\textbf{\newpage \hfill \newpage 1. Введение.} Развитие стандартов частоты и улучшение их характеристик сохраняет высокую актуальность как для  фундаментальных исследований, так и для практических применений. С использованием стандартов частоты на ионах и атомах выполнялись исследования возможного дрейфа фундаментальных постоянных~\cite{huntemann2014improved,godun2014frequency}, поиск нарушения Лоренц-инвариантности~\cite{pruttivarasin2015michelson,dzuba2016strongly}, проверка эффектов теории относительности~\cite{delva2017test}. Используя зависимость частоты атомных стандартов от гравитационного потенциала, разрабатываются высокочувствительные квантовые сенсоры. Так, относительная нестабильность на уровне $10^{-18}$ позволяет определять разность высот менее 1\,см на поверхности Земли~\cite{mehlstaubler2018atomic}. Одним из актуальных вопросов является повышение стабильности стандартов частоты, как кратковременной, так и долговременной, поскольку это существенным образом снижает время усреднения сигнала. 

Локализация одиночных ионов в ловушке Пауля позволяет  контролировать влияние внешних факторов на энергетическую структуру частиц~\cite{chou2010frequency, huntemann2016single, brewer2019al} и за счет этого достигать высокой точности измерения. Ионные стандарты обладают высокой компактностью, относительной простотой управления и малыми систематическими сдвигами частоты. Определенным недостатком стандартов частоты на одиночных ионах является более низкая относительная стабильность~\cite{itano1993quantum,peik2005laser} по сравнению с часами на нейтральных атомах при том же времени усреднения. Это обусловлено тем, что при считывании состояния ансамбля независимых частиц, проекционный шум, который главным образом ограничивает стабильность, обратно пропорционален корню из количества частиц $N$. Для оптических часов на одиночном ионе ($N=1$) усреднение частоты занимает существенно большее время по сравнению со стандартами частоты на нейтральных атомах, где обычно используются ансамбли из нескольких тысяч независимых атомов. Хотя стандарты частоты на одиночных ионах непрерывно  совершенствуются~\cite{meiser2009prospects, millo2009ultrastable}, достижение относительной нестабильности на уровне $10^{-18}$ требует  времени усреднения около 10 суток.

В идеальном случае, нестабильность стандарта  при увеличении количества независимых опрашиваемых частиц снижается  как $1/(N\tau)^{1/2}$ ($N$ - количество частиц, $\tau$ - время усреднения), что соответствует пределу квантового проекционного шума.  Однако, при использовании большего ансамбля растет вклад систематических ошибок~\cite{burt2016jpl}, например за счет межчастичных взаимодействий и неоднородности внешних полей. В последние годы в мире ведутся работы по созданию ионных стандартов на небольших ансамблях (кулоновских ионных кристаллах с $N$ порядка 10), что позволяет контролировать  вклад систематических ошибок на требуемом  уровне~\cite{herschbach2012linear,keller2019controlling,hausser2402115in}. В этих экспериментах опрашиваемые ионы являются  независимыми частицами, и нестабильность частоты снижается как $1/(N\tau)^{1/2}$. 

Ранее было показано~\cite{bollinger1996optimal}, что при использовании запутанных состояний из $N$ частиц возможно достичь  усреднения сигнала как $1/N$, что соответствует не стандартному квантовому пределу, а  гейзенберговскому пределу. Одним из таких коллективных состояний может быть $N$-частичное состояние Гринбергера-Хорна-Цайлингера (GHZ)~\cite{giovannetti2004quantum,kok2004quantum, leibfried2004toward}. Однако здесь следует учитывать, что коллективные GHZ состояния в гораздо большей степени подвержены процессам декогеренции по сравнению с возбужденным состоянием одиночной частицы. Физические механизмы, приводящие к декогеренции состояния, могут быть как внешними для квантовой системы (например, нестабильность опрашивающего лазера), так и обусловленными свойствами самой системы (спонтанный распад возбужденного состояния частиц). Учет процессов декогеренции приводит к уменьшению оптимального времени опроса в Рамси-схеме, что при детальном рассмотрении~\cite{huelga1997improvement} сводит на нет преимущество коллективных GHZ состояний перед ансамблями независимых частиц. Тем не менее, при дальнейшем изучении вопроса было показано~\cite{kielinski2024ghz}, что в случае, когда доминирующим процессом в декогеренции коллективного состояния является спонтанный распад, возможно построение методики опроса квантовой системы, обеспечивающей преодоление стандартного квантового предела при использовании коллективных GHZ состояний. Основными внешними факторами, приводящими к декогеренции, обычно являются нестабильность часовой лазерной системы и шумы магнитного поля. На сегодняшний день время когерентности часовых лазеров достигает секунд~\cite{matei20171}, что находится на уровне или превосходит время жизни возбужденного состояния некотрых атомных систем, используемых в оптических стандартах частоты, таких как, например, ионы $\text{Ca}^+$ (1.1c) и $\text{Sr}^+$ (0.4с). В данном исследовании мы работаем с кубитами на переходе $^2S_{1/2}\ (F=0, m_F=0) \to \,^2D_{3/2}\ (F=2, m_F=0)$ в ионе $^{171}\text{Yb}^+$, время жизни которого составляет 50мс, что также приводит к режиму превалирования спонтанного распада над другими процессами декогеренции в системе и возможности преодоления стандартного квантового предела.

С развитием области квантовых вычислений на цепочках ионов в ловушках появляется все больше методов для манипулирования квантовыми состояниями ионов, в том числе для создания запутанных состояний. В данной работе мы продемонстрировали подготовку GHZ состояния на двух-восьми кубитах в цепочке из 10 ионов $^{171}\text{Yb}^+$, захваченных в линейную ловушку Пауля, для чего использовалась последовательность однокубитных и двухкубитных операций, и выполнили оценку достоверности полученных состояний.

\begin{figure}
    \centering
    \includegraphics[width=\linewidth]{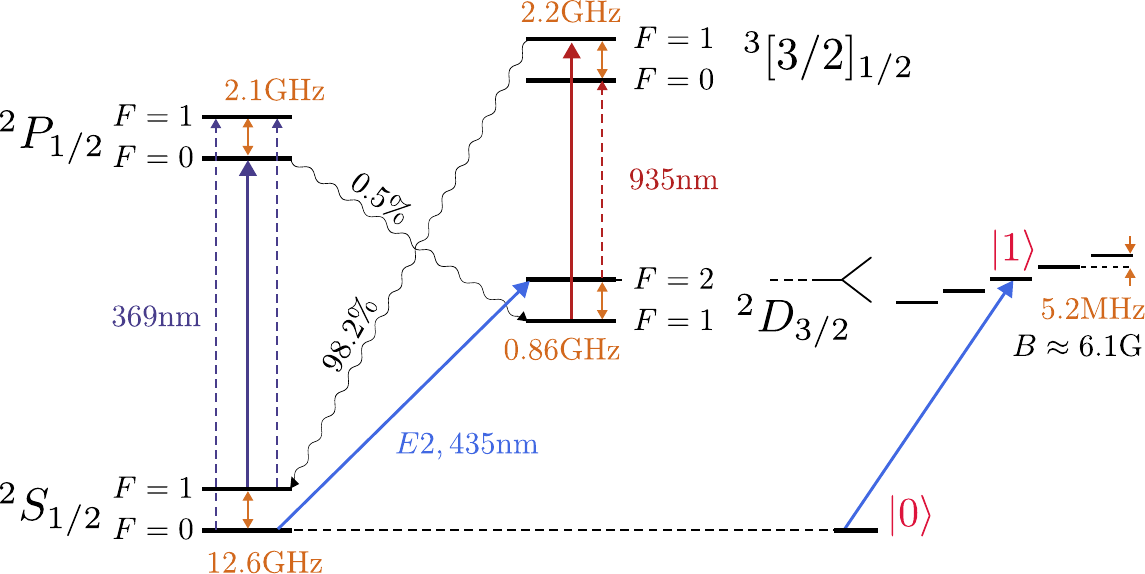}
    \caption{Рис.~\ref{fig:yb171}. Структура уровней иона $^{171}\text{Yb}^+$. Сплошными линиями показаны лазерные поля, использующиеся для лазерного охлаждения, подготовки начального состояния, выполнения операций и  считывания конечного состояния. Пунктирные линии соответствуют боковым частотам, полученным с помощью электрооптических модуляторов, которые предотвращают пленение населенности на метастабильных подуровнях сверхтонкой структуры.}
    \label{fig:yb171}
\end{figure}

\textbf{2. Экспериментальная установка.} Мы использовали ионы $^{171}\text{Yb}^+$ в качестве физической системы для демонстрации запутывания~\cite{bruzewicz2019trapped} (рис.~\ref{fig:yb171}). Ионы $^{171}\text{Yb}^+$ широко используются при создании оптических стандартов частоты,  прецизионной метрологии и спектроскопии~\cite{huntemann2016single},  а также для построения ионных квантовых процессоров. Именно на ионах $^{171}\text{Yb}^+$ были получены одни из лучших результатов для алгоритмических квантовых вычислений \cite{Quantinuum2023, wang2021single}. Лазерное охлаждение, подготовка состояний, считывание с высокой достоверностью могут быть реализованы с помощью доступных на сегодняшний день лазерных систем.

В данном ионе существует два оптических часовых перехода, которые используются в оптических стандартах частоты: E2 переход $^2S_{1/2}\ (F=0, m_F=0) \to \,^2D_{3/2}\ (F=2, m_F=0)$ и E3-переход $^2S_{1/2}\ (F=0, m_F=0) \to \,^2F_{7/2}\ (F=3, m_F=0)$. Для создания запутанности, а также  спектроскопии иона,   мы используем переход E2, соединяющий кубитные состояния: $\ket{0}=\,^2S_{1/2}\ (F=0, m_F=0)$ и $\ket{1}=\,^2D_{3/2}\ (F=2, m_F=0)$ \cite{zalivako2019improved}. Длина волны перехода составляет $\lambda=435.5~\text{нм}$, а естественная ширина линии $3~\text{Гц}$ (время жизни возбужденного состояния $53~\text{мс}$). По сравнению со  схемой кодирования кубита на компонентах сверхтонкой структуры основного состояния $^2S_{1/2}$ (частота $12.6~\text{ГГц}$), наш метод обеспечивает индивидуальную адресацию ионов в цепочке с использованием широкодиапазонных акустооптических дефлекторов (АОД) на основе кристалла $\text{TeO}_2$. 

Для захвата ионов используется линейная ловушка Пауля, помещенная в вакуумную камеру из нержавеющей стали, где вакуум после высокотемпературного отжига поддерживался ион-геттерным насосом на уровне лучше, чем $10^{-10}~\text{мбар}$. Ловушка состоит из четырех покрытых золотом медных электродов в виде лезвий, обеспечивающих радиальное удержание, и двух цилиндрических торцевых электродов для удержания вдоль оси ловушки. Изоляторы между электродами выполнены из алюмооксидной керамики. Расстояние между осью ловушки и электродами-лезвиями составило $r_0=250~\text{мкм}$. Более подробно конструкция ловушки описана в работе \cite{zalivako16qubits}.

Помимо радиочастотного сигнала, требуемого для удержания ионов, на каждый из электродов независимо прикладывалось постоянное напряжение. Это позволяет  компенсировать избыточные микродвижения ионов в ловушке и внести асимметрию между радиальными направлениями в ловушке ($x$ и $y$) для разведения колебательных мод. Амплитуда переменного напряжения, подаваемого на электроды ловушки, стабилизирована петлей обратной связи~\cite{johnson2016active}. В работе использовались следующие значения секулярных частот: $\{ \omega_x, \omega_y, \omega_z\} = 2\pi \times \{ 3.7, 3.8, 0.116 \} \text{МГц}$ для одиночного иона. 

\begin{figure}
    \centering
    \includegraphics[width=\linewidth]{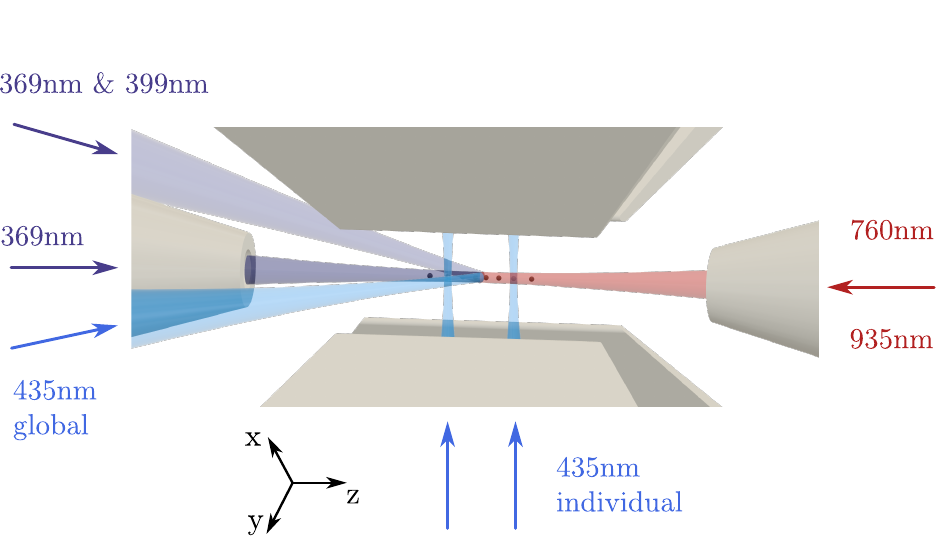}
    \caption{Рис.~\ref{fig:beams}. Схема завода лазерных пучков в ловушку Пауля. Показаны пучки охлаждения и фотоионизации (369нм \& 399нм), основной пучок доплеровского охлаждения (369нм), пучки перекачки (935нм и 760нм), а также пучки адресации, направленные перпендикулярно  оси $z$.}
    \label{fig:beams}
\end{figure}

Доплеровское охлаждение ионов проводилось на квазициклическом переходе $^2S_{1/2}(F=1)\rightarrow \, ^2P_{1/2}(F=0)$ с длиной волны 369~нм и естественной шириной линии $\Gamma=2\pi \times 20\,\text{МГц}$. В процессе охлаждения излучение лазера модулировалось по фазе электрооптическим модулятором (ЭОМ) с частотой 14.7~ГГц для перекачки населенности из метастабильного состояния $^2S_{1/2}(F=0)$ через переход $^2S_{1/2}(F=0)\rightarrow \, ^2P_{1/2}(F=1)$. Основной охлаждающий лазерный пучок был направлен под углом ко всем трем осям ловушки, в то время как еще один пучок на той же длине волны был заведен вдоль оси ловушки и участвовал в считывании, вызывая зависящую от состояния иона флюоресценцию (рис.~\ref{fig:beams}). Поскольку состояние $^2P_{1/2}(F=0)$ может с вероятностью $0.5\%$ распасться в метастабильное состояние $^2D_{3/2}(F=1)$, дополнительный лазер перекачки с длиной волны 935~нм использовался для возврата ионов из этого состояния в цикл охлаждения. Излучение этого лазера модулировалось ЭОМом на частоте 3.07~ГГц для вывода ионов из $^2D_{3/2}(F=2)$ и сброса состояний кубитов в конце каждого экспериментального цикла. Оба лазера были стабилизированы по измерителю длин волн со встроенным пропорционально-интегрально-дифференциальным регулятором.  Волномер регулярно калибровался по часовому лазеру на длине волны 871~нм~\cite{Zalivako2020, galstyan2025injection}. Также использовался лазер с длиной волны 760~нм для возврата ионов из долгоживущего состояния $^2F_{7/2}$, в которое они могут попасть из-за взаимодействия с остаточными газами в вакуумной камере.

Процесс доплеровского охлаждения проводился перед каждой экспериментальной цепочкой операций и занимал 6~мс. Температура ионов после цикла охлаждения измерялась методом анализа затуханий Раби осцилляций~\cite{semenin2022determination} и составляла 1.6~мК. После этого проводилась инициализация начального состояния $\ket{0}=\,^2S_{1/2}(F=0,m_F=0)$ путем выключения ЭОМа на 14.7~ГГц и включения ЭОМа на 2.1~ГГц в основном охлаждающем пучке лазера на длине волны 369~нм. Для проведения двухкубитных операций доплеровское охлаждение сопровождалось охлаждением до основного колебательного состояния методом охлаждения на разрешенных боковых колебательных линях перехода $\ket{0}\rightarrow \ket{1}$~\cite{Monroe1995}. После охлаждения  среднее число фононов в охлаждаемых колебательных модах составило менее 0.1.

Лазер адресации кубитов, одновременно являющийся лазером для спектроскопии часового перехода с длиной волны 435~нм является крайне важной частью лазерной системы. К нему предъявляются  высокие требования в части краткосрочной и долгосрочной стабильности частоты несущей, низкого уровня фазовых шумов, высокого  качества пучка и  позиционирования в пространстве. Наша адресующая система основана на диодном лазере с внешним резонатором на длине волны 871~нм, который методом Паунда-Древера-Холла~\cite{Drever1983} привязан к ультрастабильному резонатору с телом из стекла с  низким коэффициентом теплового расширения (ULE). Лазерная система обладает нестабильностью частоты менее $3\times10^{-15}$ на временах усреднения от 0.5~с до 50~с~\cite{Zalivako2020}. Для подавления быстрых фазовых шумов лазера в эксперименте используется прошедшее излучение через резонатор, усиленное при помощи еще одного лазерного диода методом инжекционной привязки~\cite{galstyan2025injection}. Затем излучение  лазера проходит через полупроводниковый усилитель, и затем его частота удваивается в нелинейном кристалле в резонаторе. Максимальная доступная мощность излучения на выходе системы составила 1.3~Вт. Полученный пучок на длине волны 435.5~нм делится на три пучка, в каждом из которых было установлено по акустооптическому модулятору (АОМ) для контроля частоты, фазы и мощности излучения. Два из этих пучков затем проходили через акустооптические дефлекторы (АОДы) и использовались для адресации отдельных ионов. Третий из пучков использовался для глобальной адресации всех ионов и был заведен в ловушку перпендикулярно пучкам индивидуальной адресации.

Для считывания квантового состояния ионов на 1~мс включался лазер на длине волны 369~нм, заведенный вдоль оси ловушки и модулированный на частоте 14.7~ГГц, а также лазер перекачки 935~нм без модуляции. Если состояние кубита при измерении проецируется в $\ket{0}$, то будет наблюдаться люминесценция иона, или <<светлое>> состояние. В противном случае принято говорить о <<темном>> состоянии, соответствующему состоянию $\ket{1}$. Сравнивая количество зарегистрированных фотонов с заранее откалиброванным пороговым значением, мы делаем вывод о регистрации <<светлого>> или <<темного>> состояния~\cite{semenin2021optimization}. Для сбора фотонов использовалась высокоапертурная оптическая система с $NA=0.48$ и суммарным увеличением $\times 71$. Каждый ион в плоскости изображений совмещался с торцом многомодового волокна, а другие концы волокон подводились к регистрирующим площадкам многоканального ФЭУ. Импульсы от ФЭУ в свою очередь усиливались и считывались.

\textbf{3. Однокубитные и двухкубитные операции.} Экспериментальная установка поддерживает три вида квантовых операций, составляющих универсальный набор (по которым можно разложить любой квантовый алгоритм на соответствующем количестве кубитов). Первая из них, это операция вращения состояния на сфере Блоха вокруг произвольной оси в экваториальной плоскости:
\begin{equation}
R_\phi (\theta)=\exp{-i\sigma_\phi \theta/2}, 
\label{eq:gate_one}    
\end{equation}
где $\theta$ - угол вращения, $\sigma_\phi=\sigma_x\cos{\phi}+\sigma_y\sin{\phi}$, $\sigma_x, \sigma_y$ - стандартные матрицы Паули, действующие в пространстве, образованном базисными векторами $\ket{0}$ и $\ket{1}$. Для краткости определим частные случаи этих операций $R_x(\theta)\coloneqq R_{\phi=0}(\theta)$, $R_y\coloneqq R_{\phi=\pi/2}(\theta)$.

Эта операция выполняется путем подачи лазерного импульса, резонансного для перехода $\ket{0}\xrightarrow{}\ket{1}$, с индивидуальной адресацией на каждый отдельный ион. При этом фаза $\phi$ определяется фазой кубита относительно фазы лазерного поля и контролируется акустооптическим модулятором. Угол вращения $\theta=\Omega\tau$ определяется частотой Раби $\Omega$ и длительностью импульса $\tau$. В нашем эксперименте длительность $\pi$-импульса составляла порядка 10~мкс. Точность данной операции, измеренная методом рандомизированного бенчмаркинга на клиффордовских гейтах~\cite{knill2008randomized}, составила $F_{SQ}=0.99946\pm 0.00006$.

Второй поддерживаемый вид операций --- это вращение вокруг вертикальной оси на сфере Блоха на угол $\theta$:

\begin{equation}
R_z (\theta)=\exp{-i\sigma_z \theta/2}, 
\label{eq:gate_rz}    
\end{equation}

Данный вентиль является виртуальным, то есть реализуется путем смещения фазы всех последующих лазерных импульсов на угол $\theta$~\cite{mckay_efficient_2017}.

Для запутывания состояний двух ионов в нашей системе используется вентиль Мёльмера-Зоренсена~\cite{sorensen1999quantum,molmer1999multiparticle,sorensen2000entanglement} на переходе $\ket{0}\xrightarrow{}\ket{1}$, который определяется следующим образом:

\begin{equation}
XX(\chi)=\exp{-i \frac{\chi}{2}(\sigma_x \otimes \mathbb{I}+\mathbb{I} \otimes \sigma_x)^2}. 
\label{eq:gate_two}    
\end{equation}

Эта операция является полностью запутывающей при $\chi=\pi/4$. Для ее проведения на целевую пару ионов подавалось бихроматическое лазерное излучение двумя пучками индивидуальной адресации. На каждой паре ионов были оптимизированы параметры вентиля для минимизации времени операции и увеличения стойкости к флуктуациям при значении $\chi=\pi/4$, которое наиболее часто используется в квантовых алгоритмах.

Средняя достоверность двухкубитной операции по всему регистру составила $F=0.963\pm0.005$ с учетом коррекции ошибок измерений и подготовки начального состояния.

\textbf{4. Подготовка GHZ состояний и оценка достоверности.} Мы реализовали подготовку GHZ состояний с использованием однокубитных и двухкубитных операций на цепочке ионов. Отметим, что подготовка GHZ состояний является хорошей оценкой достоверности запутывающих операций и процессов декогеренции в системе, поскольку распад таких состояний происходит существенно быстрее по сравнению с одиночными кубитами. Мы рассматривали подготовку следующих $N$-кубитных GHZ состояний, $N\in \{ 2,3,4,5,6,7,8\}$:
\begin{equation}
\ket{GHZ_N} = \frac{1}{\sqrt{2}}(\ket{0}^{\otimes N}+(-1)^{\lfloor (N-1)/2\rfloor}\ket{1}^{\otimes N}). 
\label{eq:GHZ_q}    
\end{equation}

\begin{figure}[h!]
\begin{minipage}[h]{0.9\linewidth}
\center{\includegraphics[width=0.9\linewidth]{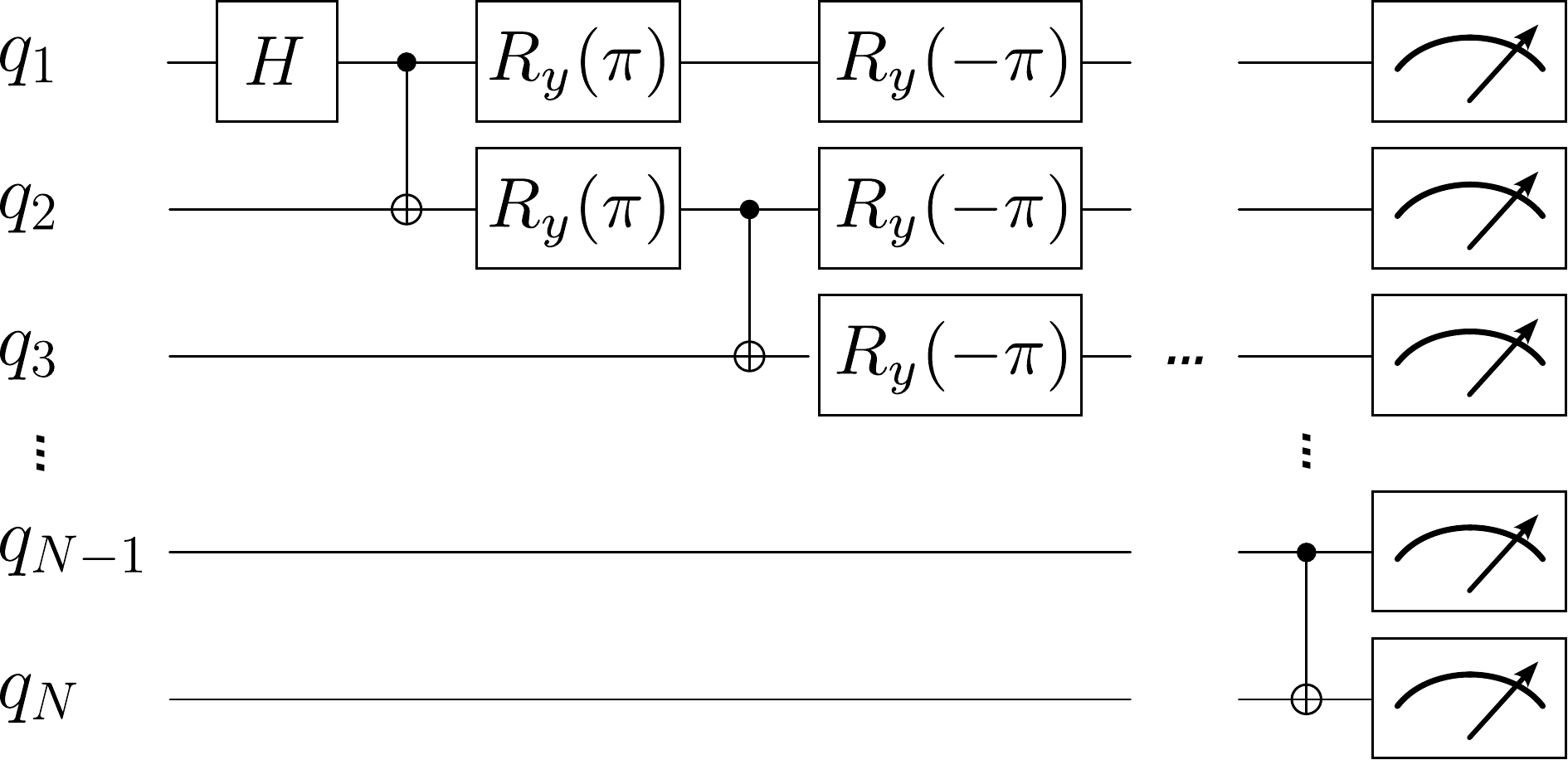}} \\(a)
\end{minipage}
\vfill
\medskip
\begin{minipage}[h]{0.9\linewidth}
\center{\includegraphics[width=1\linewidth]{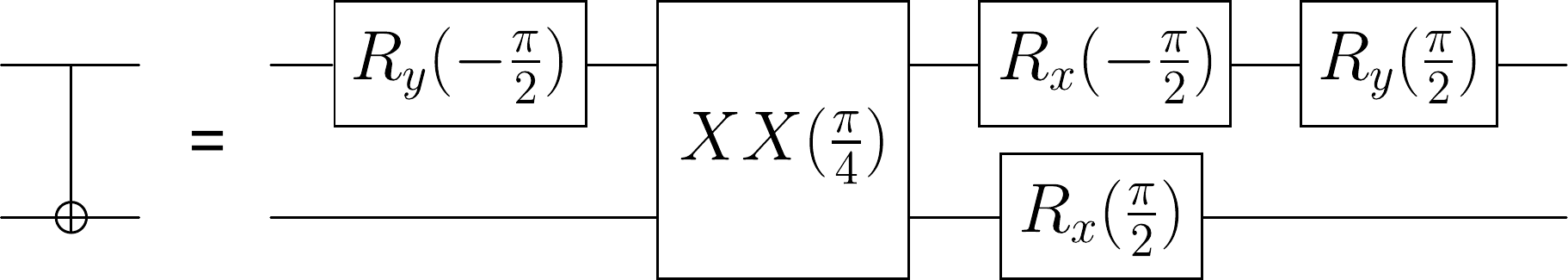}} \\(b)
\end{minipage}
\caption{Рис.~\ref{fig:preparation_scheme}. Подготовка N-кубитного GHZ состояния. Цепочка вентилей для подготовки состояния (a) и транспиляция $CX$ вентиля в набор нативных вентилей нашей системы (b).}
\label{fig:preparation_scheme}
\end{figure}

Для создания такого состояния мы использовали начальное состояние $\ket{0\cdots0}$ и применяли к нему цепочку вентилей, представленную на рис.~\ref{fig:preparation_scheme}(а), которая содержит один вентиль Адамара на первом кубите и последовательность из $N-1$ двухкубитных $CX$ вентилей таких, что $CX\ket{x,y}\mapsto\ket{x,x\oplus y}, x,y\in \{ 0,1\}$, между которыми поочередно следуют операции $R_y(\pi)$ и $R_y(-\pi)$ на всех задействованных до этого момента кубитах.  Операции $R_y(\pi)$ и $R_y(-\pi)$ являются формой импульсного динамического декаплинга, который подавляет влияние низкочастотных флуктуаций частоты лазера на точность подготовки перепутанного состояния. Отметим, что использование на этапе подготовки декаплинга не противоречит цели использования перепутанного состояния для детектирования отклонения частоты часового лазера относительно перехода. Это обусловлено тем, что чувствительность к частоте лазера в схеме опроса запутанного ансамбля частиц требуется только в период свободной эволюции созданного состояния и не требуется в процессе его подготовки. Вентиль Адамара транспилировался в последовательность вентилей $R_y(-\pi/2)$ и $R_z(\pi)$. Использованная декомпозиция двухкубитной операции $CX$ на однокубитные вращения и вентиль Мельмера-Соренсена представлена на рис.~\ref{fig:preparation_scheme}(б).

\begin{figure}
    \centering
    \includegraphics[width=\linewidth]{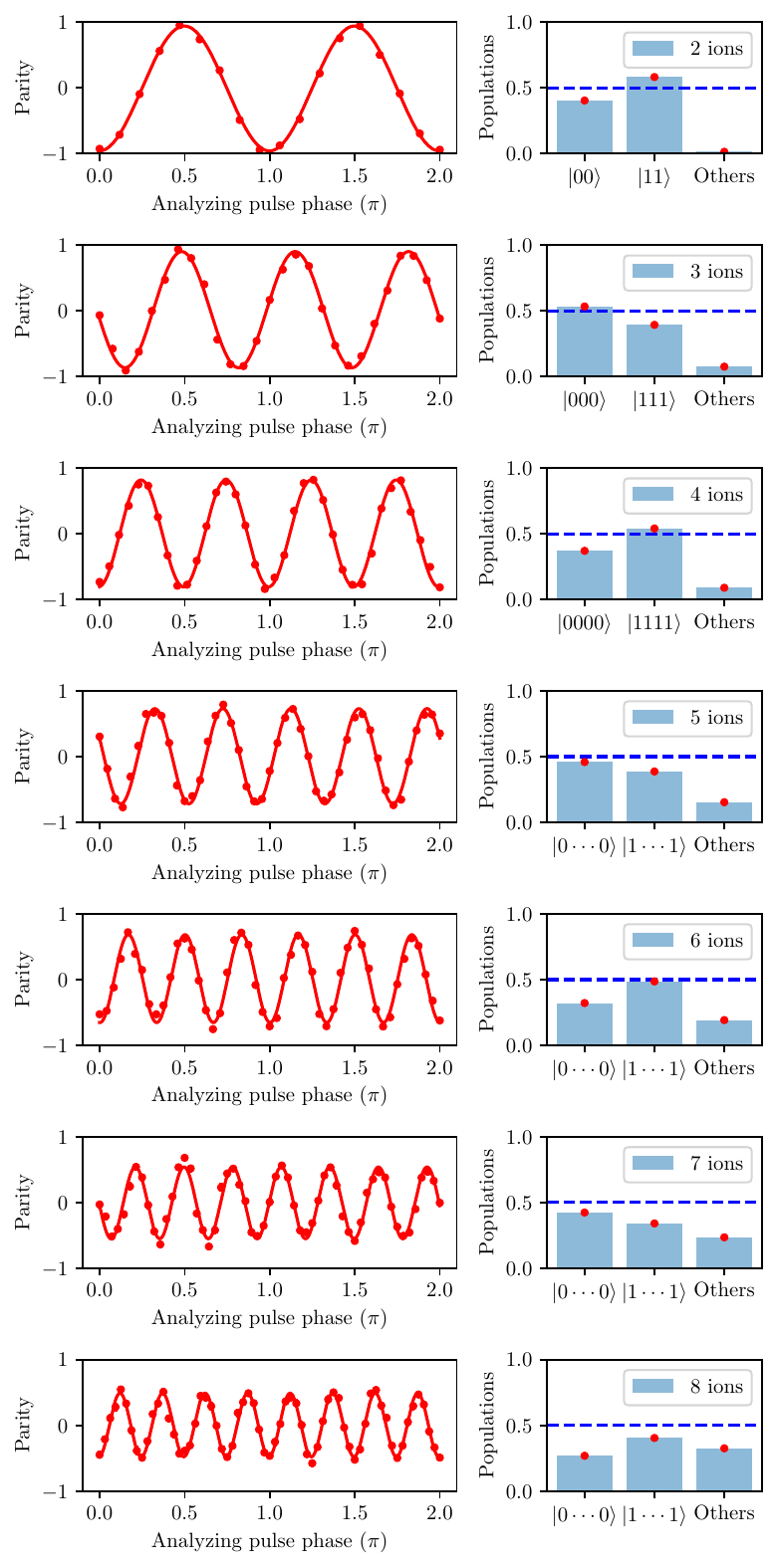}
    \caption{Рис.~\ref{fig:parity}. Для подготовленных $N$-частичных GHZ состояний измеренные значения населенностей в состояниях $\ket{0\cdots0}$ и $\ket{1\cdots1}$ (справа) и осцилляции четности после приложения ко всем ионам анализирующего $\pi/2$ импульса с варьируемой фазой $\phi$ (слева) для $N=2,3,4,5,6,7,8$. Сплошная линия на графиках четности показывает результат аппроксимации данных выражением $P(\phi)=B\times \cos{(N\phi+\phi_0)}$.}
    \label{fig:parity}
\end{figure}

Для характеризации полученных состояний $\Psi_N$ была проведена оценка достоверности $F=|\langle \Psi_N | GHZ_N \rangle|^2$ согласно методике, изложенной в работах~\cite{monz201114, leibfried2005creation}. Данный подход подразумевает проведение двух экспериментов, которые позволяют оценить диагональные и недиагональные элементы матрицы плотности, соответственно. Первый из этих экспериментов подразумевает непосредственно подготовку GHZ состояния и последующее измерение регистра в вычислительном базисе. Измеряется суммарная населенность (рис.~\ref{fig:parity}, справа) в подпространстве $\text{Span}(\ket{0}^{\otimes N}, \ket{1}^{\otimes N})$, которая соответствует сумме двух диагональных элементов матрицы плотности $\rho$: $A=\rho_{0\cdots0,0\cdots0}+\rho_{1\cdots1,1\cdots1}$. При идеальной подготовке GHZ состояния $A=1$, в то время как в случае утечки части населенности в какое-либо другое состояние значение $A$ будет уменьшаться. 

Второй эксперимент подразумевает приложение после подготовки состояния анализирующего $\pi/2$-импульса с варьируемой фазой $\phi$ ко всем задействованным ионам и измерение четности полученного состояния (разность между полной населенностью в состояниях с четным числом возбужденных частиц и полной населенностью состояний с нечетным числом возбужденных частиц). Аппроксимировав полученную зависимость четности от фазы $\phi$ функцией $P(\phi)=B\times \cos{(N\phi+\phi_0)}$, был получена амплитуда осцилляций четности $B$ (рис.~\ref{fig:parity}), которая равна сумме модулей двух недиагональных элементов матрицы плотности~\cite{monz201114}: $B=|\rho_{0\cdots0,1\cdots1}|+|\rho_{1\cdots1,0\cdots0}|$. Полная достоверность подготовки GHZ состояния дается выражением $F_{GHZ}=A/2+B/2=1/2(\rho_{0\cdots0,0\cdots0}+\rho_{1\cdots1,1\cdots1})+|\rho_{0\cdots0,1\cdots1}|$. Полученные значения достоверности представлены в таблице~\ref{tabular:fidelities}, все данные получены со SPAM-коррекцией ошибок при подготовке и считывании состояний. 

\begin{table}[h]
\caption{Таблица~\ref{tabular:fidelities}. Достоверность полученных GHZ состояний для разного числа кубит~N.}
\label{tabular:fidelities}
\begin{center}
\begin{tabular}{l|c}
$N$ & Достоверность \\
\hline
\hline
2 & 0.968(5) \\
3 & 0.904(5) \\
4 & 0.862(5) \\
5 & 0.785(6) \\
6 & 0.738(6) \\
7 & 0.655(6) \\
8 & 0.579(6) \\
\end{tabular}
\end{center}
\end{table}

Значение величины достоверности $F_{GHZ}$ напрямую связано с сепарабельностью состояния. Для проверки истинной перепутанности (несепарабельности) состояния могут быть применены операторы-свидетели. Подготовленные состояния можно проверить на истинную перепутанность с помощью оператора-свидетеля, выраженного через проектор на идеальное GHZ состояние~\cite{leibfried2005creation}:

\begin{equation}
W= \mathbb{I}-2\ket{GHZ}\bra{GHZ}.
\label{eq:GHZ_q}    
\end{equation}

Состояние является несепарабельным, если ожидаемое значение $\langle W \rangle$ для данного состояния отрицательно. Выражая его через элементы матрицы плотности, можно записать: $\langle W \rangle = 1-2F_{GHZ}$. Поэтому условие $F_{GHZ}>0.5$ является достаточным для несепарабельности состояния, что позволяет говорить о его истинной перепутанности. Строго говоря, состояния, для которых $\langle W \rangle \geqslant0~(F<0.5)$,  также могут быть истинно перепутанными, однако требуют отдельного и более сложного исследования.

\textbf{5. Заключение.} В работе была экспериментально продемонстрирована подготовка $N$-частичных GHZ состояний для $N=2,3,4,5,6,7,8$ с использованием оптических кубитов, закодированных в квадрупольном переходе на длине волны 435.5~нм в ионах $^{171}\text{Yb}^+$. Величина достоверности полученных состояний удовлетворяют критерию их истинной перепутанности. 

Важно отметить, что  измерение четности в зависимости от фазы анализирующего импульса является обобщением Рэмси-эксперимента на случай ансамбля из $N$ запутанных частиц~\cite{bollinger1996optimal}. При этом, как можно видеть из рисунка~\ref{fig:parity}, чувствительность изменения четности к изменению фазы часового лазера оказывается в $N$ раз больше, чем в случае Рэмси-спектроскопии одной частицы. Действительно,  за один период изменения фазы анализирующего импульса четность многочастичного состояния  изменяется на N периодов.

Полученный результат является важным шагом на пути создания  оптического стандарта частоты на ансамбле из запутанных ионов иттербия, что позволит преодолеть стандартный квантовый предел шумов и кратно уменьшить время усреднения частоты, хотя и потребует применения специальных методик измерений~\cite{kielinski2024ghz} в условиях декогеренции коллективных состояний из-за нестабильности часового лазера и спонтанного распада. 

\textbf{Благодарности.}
Работа выполнена при поддержке Росатома в рамках выполнения Дорожной карты "Квантовые вычисления", договор № 1.3-15/15-2021 от 5 октября 2021. Схема динамического декаплинга для приготовления GHZ-состояний была разработана Хабаровой К.Ю. при поддержке гранта РНФ № 24-12-00415.

\bibliographystyle{ieeetr}
\bibliography{biblio.bib}

\end{document}